\documentclass[twocolumn,english,aps,pra,showpacs]{revtex4}
\usepackage{amssymb}
\usepackage{amsmath}
\usepackage{graphicx}
\usepackage{epstopdf}
\usepackage{bm}
\usepackage{braket} 
\usepackage{color}
\usepackage{rotating,booktabs}

\begin{document}
\title{Bose-Fermi Mapping and Multi-Branch Spin Chain Model for Strongly Interacting Quantum Gases in One-Dimension: Dynamics and Collective Excitations}
\author{Li Yang$^1$, and Han Pu$^{1,2}$}

\affiliation{$^{1}$Department of Physics and Astronomy, and Rice Center for Quantum Materials,
Rice University, Houston, TX 77251, USA \\
$^2$Center for Cold Atom Physics, Chinese Academy of Sciences, Wuhan 430071, P. R. China}

\begin{abstract}
We show that the wave function in one spatial sector $x_1<x_2<\cdots < x_N$ (with $x_i$ being the coordinate of $i$th particle) of a one dimensional spinor gas with contact $s$-wave interaction, either bosonic or fermionic, can be mapped to the direct product of the wave function of a spinless Fermi gas with short-range $p$-wave interaction and that of a spin system governed by spin parity projection operators. Applying this mapping to strongly interacting spinor gases, we obtain a generalized spin chain model that captures both the static and dynamics properties of the system. Using this spin chain model, we investigate the breathing mode frequency and the quench dynamics of strongly interacting harmonically trapped spinor gases. 
\end{abstract}


\maketitle

\section{Introduction}
Strongly interacting many-body systems exhibit rich physics, but in general pose as tremendous theoretical challenges. Under certain circumstances, a system can be mapped into another one much more amenable to theoretical study. 
The Bose-Fermi mapping is one such example \cite{Girardeau1960,BFreview}. It maps a system of one dimensional (1D) spinless bosons with infinite repulsive two-body contact interaction to a system of spinless non-interacting fermions. This mapping is based on the idea that, due to the infinite interaction, the relative wave function between two identical bosons must vanish when $x_i=x_j$, which mimics the quantum statistics between two identical fermions. This mapping was later generalized by Cheon and Shegehara \cite{Cheon1998,Cheon1999} who mapped a system of spinless bosons with $s$-wave contact interaction characterized by strength $g$
\begin{equation}
V_s = g\,\sum_{i<j} \delta(x_{ij}) \,,\;\;\;x_{ij} \equiv x_i-x_j\,,\label{b}
\end{equation}  
to a system of spinless fermions interacting with each other via a short-range $p$-wave interaction of strengh $1/g$, whose pseudo-potential form can be written as \cite{Sen1999,Sen2003,Grosse2004,Hao2007,regularization}
\begin{equation}
V_p = -\frac{4}{g}\,\sum_{i<j}  \overleftarrow{\partial}_{x_{ij}}\delta(x_{ij})\overrightarrow {\partial}_{x_{ij}^{-}}\,. \label{f}
\end{equation}

Much richer physics can be obtained if the particles possess spin degrees of freedom. The goal of the current work is to present a general mapping that works for a 1D quantum gas with arbitrary spin. Applying this mapping to a strongly $s$-wave interacting spinor quantum gas, we show that we can construct an effective spin chain model that accurately captures both the static and dyanmic properties of the system. 1D cold atomic systems have been realized in experiments by strongly confining the atoms along two transverse directions such that the transverse dynamics is frozen into the single-particle ground state. For two recent reviews, see Refs.~\cite{rmp1,rmp2}.

The paper is organized as follows. In Sec. II, we present the generalized mapping, in which the wave function of a 1D $s$-wave interacting spinor quantum gas in one spatial sector is mapped to a direct product of the wave function of a spinless fermions with $p$-wave interaction and that of a spin system. In Sec. III, we apply this mapping to a system of strongly interacting spinor quantum gas and construct effective multi-branch spin chain models. In Sec. IV, we further consider the spin chain model for a harmonically trapped system, and show that the multi-branch spin chain Hamiltonian leads to an efficient way of calculating the breathing mode frequency. More details of the breathing mode frequency of strongly interaction Bose and Fermi gases are presented in Sec. V. In Sec. VI, we investigate the quench dynamics using the multi-branch spin chain model, demonstrating its utility in a dynamical situation. Concluding remarks are presented in Sec. VII. Some technical details are presented in the three Appendices.

\section{Generalized Bose-Fermi mapping}
Generalizing the original Bose-Fermi mapping of Girardeau to spinor systems was first proposed by Girardeau and Olshanii \cite{Girardeau2004_1,Girardeau2004_2} who showed that 1D spinor Fermi gas and Bose gas can be mapped into each other, where the even-wave interaction (e.g., $V_s$) in one is mapped to the odd-wave interaction (e.g., $V_p$) in the other. This mapping can be understood as follows: The even relative spatial wave function under $V_s$ and the odd relative spatial wave function under $V_p$ satisfy exactly the same boundary condition:
$ \lim_{x_{ij} \rightarrow 0^+} g \psi (x_{ij}) = 2 \psi'(x_{ij}) \,.$

Motivated by these past works, here we present a different, but related, mapping as follows: A 1D spinor gas, either bosonic or fermionic, interacting with contact $s$-wave two-body interaction, governed by Hamiltonian
\begin{equation}
H =H_0+V_s= \sum_{i=1}^N \left[ -\partial_i^2/2 + V(x_i) \right] +V_s \,,\label{h}
\end{equation}
where $H_0$ is the single-body Hamiltonian with $V$ representing the external trapping potential, can be mapped to the direct product of a spinless $p$-wave interacting Fermi gas and a spin chain system under the Hamiltonian 
\begin{equation}
H_F=H_0+H_p\,, \label{hf}
\end{equation} 
where 
\begin{equation}
H_{p}=- \frac{4N!}{g}\,\sum_{i=1}^{N-1}\overleftarrow{\partial}_i\delta(x_i-x_{i+1})\theta^1\overrightarrow{\partial}_{i-}
\otimes \hat{P}^{s,a}_i \,. \label{pesudopotential}
\end{equation}
Note that this mapping is defined in one spatial sector defined by $\theta^1$, which is the sector function (i.e., generalized Heaviside step function) of spatial coordinates, whose value is one in spatial sector $x_1<x_2<\cdots<x_{N}$, and zero otherwise. The operators $\hat{P}^{s,a}_i=(1\pm{\cal E}_{i,i+1})/2$ are spin projection operators that project out symmetric and antisymmetric spin states, respectively, where ${\cal E}_{ij}$ is the exchange operator that exchanges the $i^{\rm th}$ and $j^{\rm th}$ spins. If the original spinor gas is bosonic (fermionic), one should take $\hat{P}^s_i$ ($\hat{P}^a_i$).

To see how this mapping works, let us consider a 1D spinor quantum gas with a total of $N$ particles governed by Hamiltonian (\ref{h}). The $N$-body wave function can be written as 
\begin{eqnarray}
& \Psi (x_1,x_2,...,x_N,\sigma_1,\sigma_2,...,\sigma_N)= \nonumber \\
& \sum_P (\pm 1)^P P \left(\Psi^1(x_1,x_2,...,x_N,\sigma_1,\sigma_2,...,\sigma_N) \right)\,, \label{wf}
\end{eqnarray}
where $\sigma$'s are the spin variables, $P$ represents permutation, and $\Psi^1=\Psi\theta^1$ is the wave function in the spatial sector $\theta^1$. Equation (\ref{wf}) is a manifestation of a special property of 1D system that the spatial domain of the wave function can be separated into $N!$ disconnected subdomains labeled by various spatial orders, and the wave function in one spatial sector (say, $\Psi^1$ as defined in spatial sector $\theta^1$) has the complete information of the total wave function, as the values of the wave function in different subdomains are related by permutation operation \cite{sector}. Furthermore, in the $\theta^1$ spatial sector, the wave function $\Psi^1$ can be represented as:
\begin{eqnarray}
&\Psi^1(x_1,x_2,...,x_N,\sigma_1,\sigma_2,...,\sigma_N)= \nonumber \\
& \sum_{\alpha,\beta} A_{\alpha \beta} \,\varphi_{\alpha}(x_1,x_2,...,x_N)\chi_{\beta}(\sigma_1,\sigma_2,...,\sigma_N) \,,
\end{eqnarray}
where $A_{\alpha \beta}$ are superposition coefficients, $\varphi$'s and $\chi$'s are spatial and spin wave functions, respectively.

Now we map the wave function (\ref{wf}) in the original representation into the following one:  
\begin{equation}
\sum_{\alpha,\beta} A_{\alpha \beta} \,\sum_P (-1)^P P(\varphi_{\alpha})\otimes\chi_{\beta}\,. \label{wfmapping}
\end{equation}
The mapped system is governed by the Hamiltonian $H_F$ in Eq.~(\ref{hf}). Its spatial wave function describes a spinless $p$-wave interacting Fermi gas, while the spin wave function is the eigenstate of spin projection operator $\hat{P}^{s,a}_i$. 
This mapping is defined in the spatial sector $\theta^1$, as only in this spatial sector, the boundary conditions can be mapped into each other in the two representations. In the original representation whose wave function is represented by Eq.~(\ref{wf}), at the sector boundaries $x_i=x_{i+1}$, the parities of the spatial wave function $\varphi_\alpha$ and that of the spin wave function $\chi_\beta$ are linked as the total parity has to be odd (for fermions) or even (for bosons). In the mapped representation (\ref{wfmapping}), however, this link is not present as the spatial wave function is always odd. The quantum statistics of the original system is taken care of by the spin parity project operator $\hat{P}^{s,a}_i$ in the mapped Hamiltonian $H_F$. In Appendix A, we use a simple example of two atoms to further justify the form of $H_p$ in Eq.~(\ref{pesudopotential}).

We emphasize that this mapping is exact and valid for arbitrary values of $g>0$ \cite{LijunYang2015}. However, it is particularly useful for systems with large interaction strength $g$, for which the mapped system is a weakly interacting $p$-wave spinless Fermi gas, which can be calculated perturbatively. This allows us to gain valuable insights into the original  strongly interacting systems. Furthermore, By mapping the spinor system into the direct product of a spinless fermionic system and a spin chain, the size of the Hilbert space is significantly reduced, hence efficient numerical tools can be constructed to study the system. In the following, we will present a more detailed study to showcase the application of this mapping. 
 
\section{Multi-branch spin chain model} 
Consider a trapped spinor gas with $N$ total atoms governed by Hamiltonian (\ref{h}). Although our theory is valid for arbitrary $V$, we will focus on harmonic trapping potential $V(x) = x^2/2$, which not only is the most experimentally relevant, but also possesses special symmetry properties that we will exploit later. We have adopted a dimensionless unit system where $\hbar = m =\omega=1$, with $m$ and $\omega$ being the atomic mass and the trap frequency, respectively. The interaction Hamiltonian of the mapped system is given by $H_p$ in (\ref{pesudopotential}).

For large $g$, we work on this mapped system, and treat $H_p$ as a perturbation to the single-body Hamiltonian $H_0$. 
The unperturbed system is simply an ideal Fermi gas, whose ground state is formed by putting one atom in each of the lowest $N$ single-body states, as schematically shown in Fig.~\ref{Fig1}(a), with energy $E^{(0)}=N^2/2$, and the ground state wave function is a Slater determinant which we denote as $\varphi_0$. In the context of the original spinor system, this corresponds to the Tonks-Girardeau (TG) limit with $g=\infty$, for which the ground state possess spin degeneracy as its energy is completely independent of the spin configuration. For large but finite $g$ \cite{note}, to first order in $H_p$ (i.e., in $1/g$), we can readily derive an effective Hamiltonian:  
\begin{equation}
H_{\rm sc}^{(0)}=E^{(0)}+\langle \varphi_0|H_p|\varphi_0 \rangle = E^{(0)}-\frac{1}{g} \,\sum_{i=1}^{N-1} C_i^{(0)}\, (1 \pm {\cal E}_{i,i+1}) \,, \label{heff}
\end{equation}
with the coefficients $C_i^{(0)}$ given by 
\begin{equation}
{C}_i^{(0)}=2N!\int dx_1...dx_N \,|\partial_i\varphi_0|^2\,\delta(x_i-x_{i+1})\theta^1 \,.\label{ci}
\end{equation}
This is exactly the inhomogeneous spin chain Hamiltonian for a 1D strongly interacting quantum gas recently derived  by several groups using different methods \cite{Deuretzbacher2014,Volosniev2014,Yang2015}.

\begin{figure}[h]
\includegraphics[width=7cm]{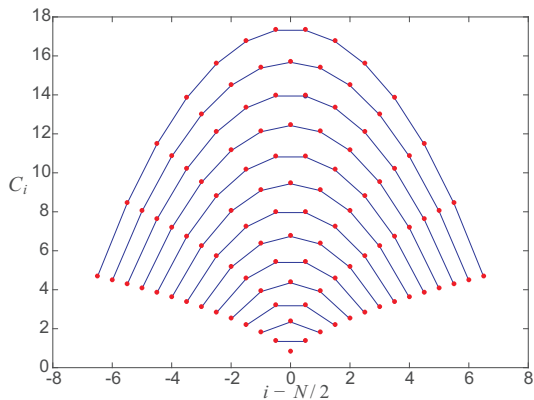}
\caption{(color online) The red dots represent results from the LDA approximation for $C_{i}^{(0)}$ using Eqs.~(\ref{CiLDA1}) and (\ref{alphaEq}). The blue solid lines are exact integral of $C_{i}^{(0)}$ by a method similar to calculating $\rho^{(i)}(z)$ in \cite{sector}. The comparison are for particle numbers run from 2 (bottom) to 15 (top). }\label{compLDA}
\end{figure}

An $N$-dimensional integral is involved in evaluating the local exchange coefficients $C_i^{(0)}$. A numerically efficient way of calculating $C_i^{(0)}$ has recently been provided in Ref.~\cite{local, local1}. Here, we develop a local density approximation (LDA) method to calculate $C_i^{(0)}$ for particles in a harmonic trap in a semi-analytical way. Under the LDA, $C_i^{(0)}$ are approximated as \cite{Deuretzbacher2014,Matveev2004,Guan2007,Matveev2008}
\begin{equation}
C_{i}^{(0)}=\frac{\pi^{2}}{3}n_{\rm TG}^{3}(y_{i}),\;\; i=1,2,...,N-1 \,, \label{CiLDA}
\end{equation}
where $n_{\rm TG}(x)=\frac{1}{\pi}\sqrt{2N-x^{2}}$ is the Tonks-Girardeau density profile which is the same as the density profile of a spinless Fermi gas, and $y_i$ are defined as 
\begin{equation}
\int_{-\sqrt{2N}}^{y_{i}}dx\;n_{\rm TG}(x)=i \,,\label{boundary}
\end{equation}
which is the average boundary of the $i$th and $(i+1)$th particle. Equations.~(\ref{CiLDA}) and (\ref{boundary}) are equivalent to
\begin{equation}
C_{i}^{(0)}=\frac{1}{3\pi}\left(2N\right)^{3/2}{\rm sin}^{3} \left(\frac{\alpha_{i}}{2} \right)\,, \label{CiLDA1}
\end{equation}
where $\alpha_i$ is the solution of equation 
\begin{equation}
\alpha_{i}-2\pi\frac{i}{N}={\rm sin}(\alpha_{i}) \,.\label{alphaEq}
\end{equation}
The comparison with the exact $C_i^{(0)}$ calculated by a similar method as calculating $\rho^{(i)}(z)$ in \cite{sector} is shown in Fig.~\ref{compLDA}, from which we see that even for very few particles the LDA results agree with the exact values very well.

\begin{figure}[h]
\includegraphics[width=6cm]{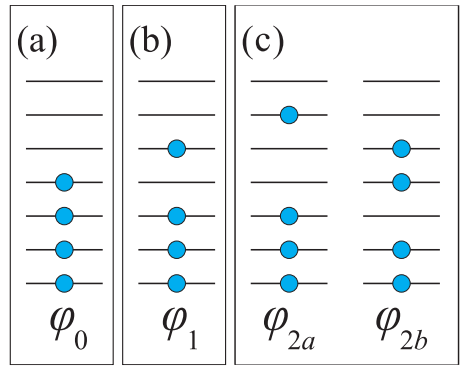}
\caption{(color online) Schematic representation of the ground state (a), the first excited state (b), and the second excited states (c) of a harmonically trapped ideal spinless Fermi gas. }\label{Fig1}
\end{figure}

Previous works have established that the spin chain model represented by Hamiltonian (\ref{heff}) describes rather accurately the ground state properties of the strongly interacting spinor gas to order $1/g$. To provide a more complete description of the system, and in particular of the {\em dynamical} properties of the system which has largely been neglected in previous works \cite{dyn}, we now extend the calculation to include excited eigenstates of the unperturbed Hamiltonian $H_0$ to construct a multi-branch spin chain model. 

The excited eigenstates of the non-interacting system governed by $H_0$ can be easily constructed. The first excited state, with wave function denoted as $\varphi_1$ and represented in Fig.~\ref{Fig1}(b), is obtained by promoting the atom at the Fermi level in the ground state $\varphi_0$ to the next single-particle state. Treating $H_p$ as a perturbation on this manifold leads to the following spin-chain Hamiltonian:
\begin{equation}
H_{\rm sc}^{(1)} = E^{(1)}- \frac{1}{g} \,\sum_{i=1}^{N-1} C_i^{(1)}\, (1 \pm {\cal E}_{i,i+1}) \,, \label{h1}
\end{equation}
where $E^{(i)}=E^{(0)}+i$ is the energy of the $i^{\rm th}$ excited state of the ideal Fermi gas, and the coefficients $C_i^{(1)}$ are given by Eq.~(\ref{ci}) with $\varphi_0$ replaced by $\varphi_1$.

The second excited state of the unperturbed system, as represented in Fig.~\ref{Fig1}(c), is two-fold degenerate with wave functions denoted as $\varphi_{2a}$ and $\varphi_{2b}$, respectively. In general, higher degree of degeneracy is expected for higher excited states. In the presence of degeneracy, the perturbation $H_p$ can in principle mix different degenerate spatial states, leading to spin-orbit coupling between the charge (i.e., spatial) and the spin sectors. When this is the case, such excited manifold cannot be described by a spin-chain Hamiltonian. 
However, in the following, we will show that the special symmetry properties of harmonic trapping potential allows us to construct the spin chain model for low-lying excited manifolds. Furthermore, the local exchange coefficients $C_i$ for these low-lying excited manifolds can be obtained from the corresponding coefficients for the ground manifold [Eq.~(\ref{ci})] without any extra calculations. This provides significant insights into the low-lying collective excitation modes for harmonically trapped spinor quantum gases.

\section{Symmetry properties of harmonic trap}
Consider first an ideal gas of harmonically trapped $N$ spinless fermions under Hamiltonian $H_0$. The center-of-mass (COM) motion can be separated from the relative internal motion. The COM dipole mode can be excited by the operator $Q^\dag=(K-iP)/\sqrt{2}$, where $K$ and $P$ correspond to the COM position and momentum operators, respectively. (For the detailed definition and properties of these operators, see Appendix B.) 
The internal states can be classified into irreducible representations of the SO(2,1) algebra obeyed by the internal operators \cite{Pitaevskii1997,Werner2006,Nishida2007,Moroz2012}. The first excited state $\varphi_1$ (Fig.~\ref{Fig1}(b)), represents the lowest COM dipole excitation and is generated from the ground state $\varphi_0$ (Fig.~\ref{Fig1}(a)) by applying $Q^\dag$ once, i.e., $\varphi_1 = Q^\dag \,\varphi_0$. Whereas the second excited manifold can be generated from $\varphi_0$ in two different ways: 
\begin{equation} \varphi_Q = (Q^\dag)^2\, \varphi_0\,,\;\;\; \varphi_B = B^\dag \,\varphi_0 \,,\label{qb}
\end{equation}  
where $\varphi_Q$ represents the second COM dipole excitation and $\varphi_B$ the first internal breathing excitation. $\varphi_Q$ and $\varphi_B$ have the same energy $E^{(2)}$ and are in fact linear superpositions of $\varphi_{2a}$ and $\varphi_{2b}$ represented by Fig.~\ref{Fig1}(c). 

For the mapping we discussed earlier, the charge degrees of freedom of a strongly interacting harmonically trapped spinor gas is mapped to a spinless Fermi gas interacting with the $p$-wave pseudo-potential $H_p$ given in  Eq.~(\ref{pesudopotential}). Since $H_p$ only affects in the internal degrees of freedom, the separation of the COM motion and internal motion discussed above remains valid. An immediate conclusion one can draw is that $H_p$ would not affect the energies of the COM dipole states generated by $Q^\dag$ as $[H_p, Q^\dag]=0$. Consequently, the COM dipole excitation frequencies are not shifted by the interaction. This is simply the manifestation of the Kohn-Sham theorem for a system of harmoincally trapped particles. A direct consequence of this is that the coefficients in the spin chain Hamiltonian $H_{\rm sc}^{(1)}$ for the first excited state (see Eq.~(\ref{h1})) are the same as the corresponding coefficients in $H_{\rm sc}^{(0)}$ for the ground state (see Eq.~(\ref{heff})), i.e., $C_i^{(1)}=C_i^{(0)}$. Hence $H_{\rm sc}^{(1)}$ and $H_{\rm sc}^{(0)}$ only differ by a constant shift of $E^{(1)}-E^{(0)}=1$, which is the frequency of the lowest COM dipole mode.

Now let us turn to the second excited manifold which contains two degenerate states $\varphi_Q$ and $\varphi_B$ defined in Eq.~(\ref{qb}). Due to the fact that $Q^\dag$ is a COM operator, and both $B^\dag$ and $H_p$ affect only the internal motion, the interaction does not couple $\varphi_Q$ and $\varphi_B$. As a result, we can write down the effective spin chain Hamiltonians for these two states separately:
\begin{eqnarray}
H^{Q,B}_{\rm sc} &=& E^{(2)}- \frac{1}{g} \,\sum_{i=1}^{N-1} {C}_i^{(Q,B)}\, (1 \pm {\cal E}_{i,i+1}) \,.
\end{eqnarray}
Furthermore, for the same reason that $C_i^{(1)}=C_i^{(0)}$, we also have $C_i^{(Q)}=C_i^{(0)}$. Quite amazingly, there also exists a simple relation between $C_i^{B}$ and $C_i^{(0)}$ which can be proved using a recursion relation for the SO(2,1) algebra \cite{Moroz2012,Levinsen2015} (for a detailed derivation, see Appendix C): 
\begin{equation}
\frac{ C_i^{B}}{C_i^{(0)}} = 1+ \frac{3}{2(N^2-1)} \,,\label{factor}
\end{equation}
which means that $H_{\rm sc}^{(B)}$ and $H_{\rm sc}^{(0)}$, apart from a constant shift of $E^{(2)}-E^{(0)}=2$, only differ by a constant factor given in Eq.~(\ref{factor}). The energy difference between the ground states of $H_{\rm sc}^{(B)}$ and $H_{\rm sc}^{(0)}$, which gives the frequency of the lowest breathing mode $\omega_B$, is therefore
\begin{equation}
\omega_B = 2 + \frac{3}{2(N^2-1)} E_g \,,\label{ob}
\end{equation}
where $E_g = \langle H_{\rm sc}^{(0)} \rangle - E^{(0)}$ is the ground state energy of the spin chain Hamiltonian $H_{\rm sc}^{(0)}$ measured with respect to $E^{(0)}$. Hence, unlike the COM dipole mode, the breathing mode frequency receives an interaction-dependent shift away from the non-interacting value of 2. In the strongly interacting regime, this shift $\delta \omega_B \equiv \omega_B-2 \propto 1/g$ and vanishes exactly in the TG limit of $g=\infty$. We note that the breathing of 1D quantum gases have been investigated in several recent experiments \cite{hans,fallani}.

\section{Breathing mode for harmonically trapped quantum gas} 
Let us now take a further look at the breathing mode, whose frequency $\omega_B$ is completely determined by the ground state energy of the spin chain Hamiltonian $H_{\rm sc}^{(0)}$. For a system of spinor Bose gas with {\em arbitrary} spin and arbitrary population distribution among spin components, it is quite obvious that the ground state of $H_{\rm sc}^{(0)}$ is obtained by arranging the atoms into a fully spin symmetric configuration such that $\langle {\cal E}_{i,i+1} \rangle =1$, and correspondingly the ground state energy is given by 
\begin{equation}
E_g^{\rm boson}= -\frac{2}{g} \sum_{i=1}^{N-1} C_i^{(0)} \,,\label{eg}
\end{equation}
which, for a given trapping potential, only depends on the total number of atoms $N$. Taking $N\rightarrow\infty$, using the LDA result Eq.~(\ref{CiLDA1}) for $C_i^{(0)}$, and converting the sum in Eq.~(\ref{eg}) into an integral:
\begin{align}
E_g^{\rm boson}
&\approx -\frac{1}{g} \frac{\left(2N\right)^{5/2}}{3\pi}\int_0^1 d\beta \, {\sin}^{3} \left[\frac{\alpha(\beta)}{2} \right] \nonumber \\
&=-\frac{1}{g}\frac{128\sqrt{2}}{45\pi^2}N^{5/2}\approx-\frac{1}{g}0.408N^{5/2} \,,\label{Egintegral} 
\end{align}
where $\beta=i/N\in(0,1)$. This result is consistent with the previous result obtained for spinless bosons near the TG limit \cite{Astrakharchik2005,Zhang2014,Paraan2010}, which gives another indication that our LDA approximation for $C_i^{(0)}$ is excellent.
Correspondingly, the interaction-induced shift of the breathing mode frequency is
\begin{equation}
\delta \omega_B^{\rm boson} = \frac{3}{2(N^2-1)} E_g^{\rm boson} \approx -\frac{1}{g} \frac{64 \sqrt{2}}{15 \pi^2} \,N^{1/2} \,.\label{do}
\end{equation}

\begin{figure}[h]
\includegraphics[width=8.5cm]{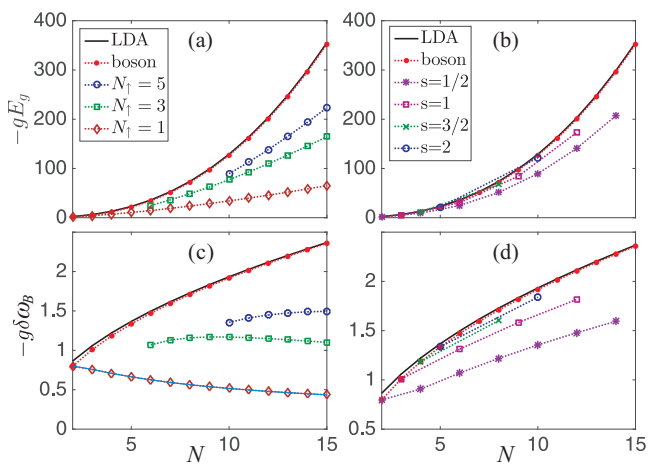}
\caption{(color online) Ground state energy (a, b) and breathing mode frequency shift (c, d) as functions of $N$. In ({a}) and ({c}), we present results for bosons and spin-$1/2$ fermions with various $N_{\uparrow}/N$. In ({b}) and ({d}), we present results for bosons, and fermions with different spin $s$ and equal population in each spin component. For bosons, the ground state energy and the breathing mode frequency shift are independent of spin. The black solid lines represent the analytic LDA results for bosons given in Eqs.~(\ref{Egintegral}) and (\ref{do}).}\label{OmegaFig}
\end{figure}

The fermionic case is more complicated.
For a spin-$s$ Fermi gas with a fully spin antisymmetric configuration, its  ground state energy is the same as in the bosonic case, given by Eq.~(\ref{eg}), as the two systems possess the same spatial wave function. This spin configuration, however, can only occur if the number of spin components $2s+1 \ge N$ and no more than 1 fermions occupy the same spin component \cite{yang}. 
In Fig.~\ref{OmegaFig}(a), we plot the spin chain ground state energy $E_g$ as functions of $N$, with the corresponding breathing mode frequency shift $\delta \omega_B$ plotted in Fig.~\ref{OmegaFig}(c). The symbols are obtained by numerically calculate the coefficients $C_i^{(0)}$ and then diagonalize the spin chain Hamiltonian $H_{\rm sc}^{(0)}$. The red dots are the results for bosons. We also plot the analytical results based on the LDA (Eqs.~(\ref{Egintegral}) and (\ref{do})) as black solid lines. As one can see, the LDA results agree very well with the numerical results even for small $N$. Other symbols in the figure correspond to $E_g$ and $\delta \omega_B$ for spin-1/2 Fermi gas with different population distribution in the two spin components. In Fig.~\ref{OmegaFig}(b) and (d), we plot respectively $E_g$ and $\delta \omega_B$ as functions of $N$ for Fermi gases with different spin $s$ and equal population in each spin component. As one can see, for fixed $N$, as $s$ increases, the fermionic results approach the bosonic ones. As $2s+1 \ge N$, the two results matches exactly. This behavior has been recently seen in the experiment \cite{fallani}.

\section{Quench dynamics} 
\begin{figure}[h]
\includegraphics[width=8.5cm]{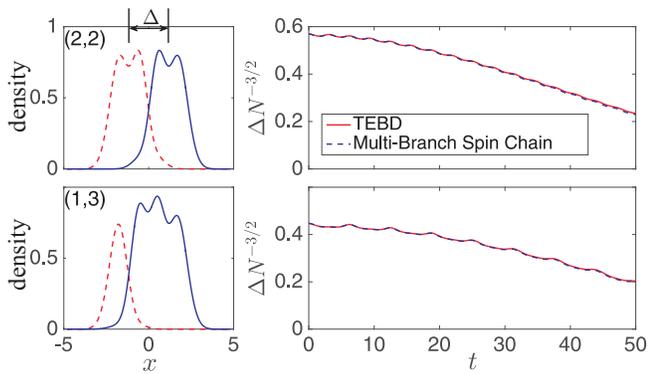}
\caption{(color online) Evolution after a sudden quench of the spin-dependent magnetic gradient for a harmonically trapped spin-1/2 Fermi gas. Upper Panel: $N_{\uparrow}=N_{\downarrow}=2$ and $g=100$. The left panel shows the spin density profiles for spin-up (dashed line) and spin-down (solid line) component before the quench. The right panel shows the after-quench evolution of $\Delta$, the center-of-mass separation of the two spin species.  The blue dashed line is obtained from the multi-branch spin chain model by including 8 excited branches. The red solid line is the TEBD result. The lower panel is the same as the upper panel except that $N_\uparrow =1$ and $N_\downarrow=3$.}\label{gradientquench}
\end{figure}

Finally, we demonstrate the application of multi-branch spin chain model to simulate the dynamics of the system. To this end, we consider a spin-1/2 Fermi gas initially prepared in a harmonic trap subject to a spin-dependent magnetic gradient that separates the COM position of the two spin components (see the left panels of Fig.~\ref{gradientquench}). In the presence of such a spin-dependent magnetic gradient, the Hamiltonian is given by 
\[ H = \sum_{i=1}^N  \left[ -\partial_i^2/2 + V(x_i) -G x_i\sigma_i^z\right] +g\,\sum_{i<j} \delta(x_{ij}) \,,\]
where $G$ is the strength of the magnetic gradient. The corresponding spin chain Hamiltonian for the ground manifold now takes the form
\begin{equation}
H_{\rm sc}^{(0)}=E^{(0)}-\frac{1}{g} \,\sum_{i=1}^{N-1} C_i^{(0)}\, (1 \pm {\cal E}_{i,i+1}) -G\sum_{i=1}^N D_i^{(0)}\sigma_i^z \,, \label{heff1}
\end{equation}
where 
\begin{equation}
D_i^{(0)}=N!\int  dx_1...dx_N \,x_i |\varphi_0|^2\theta^{1},\;\; i=1,2,...,N
\end{equation}
$D_i^{(0)}$ has a physical meaning of the average position of $i$th particle, which naturally leads to a LDA expression:
\begin{align}
D_i^{(0)}
=\int_{y_{i-1}}^{y_{i}}dx\,xn_{\rm TG}(x) 
=C_{i-1}^{(0)}-C_{i}^{(0)} \,.
\end{align}
This relation can even be numerically proven to be true for exact $C_i^{(0)}$ and $D_i^{(0)}$ for arbitrary particle numbers $N$ without envoking the LDA.

At $t=0$, the magnetic gradient is suddenly quenched to zero \cite{fm} and we plot the COM separation between the two spin components, $\Delta$, in the right panels of Fig.~\ref{gradientquench} as a function of time. The upper panel considers a situation with $N_{\uparrow}=N_{\downarrow}=2$. This situation is examined in our earlier work \cite{Yang2015} using the single-branch spin chain Hamiltonian $H_{\rm sc}^{(0)}$, benchmarked with the numerically unbiased time-evolving block decimation (TEBD) calculation. The TEBD result, which takes a few days to obtain on a laptop, is reproduced here as red solid lines. The result from the multi-branch spin chain model, which takes less than a minute to obtain, is plotted as blue dashed lines and is in perfect agreement with the TEBD result. For the short time scale we plotted, $\Delta$ decreases in time. The single-branch spin chain result (see Fig.~6 of Ref.~\cite{Yang2015}) captures this behavior, but could not produce the small-amplitude oscillations that can be clearly seen in the TEBD simulation. Analysis shows that the small-amplitude oscillation is mainly due to the coupling to the lowest breathing mode which can only be captured if the second excited manifold is included in the spin chain model. The lower panel of Fig.~\ref{gradientquench} considers a similar quench dynamics with $N_\uparrow =1$ and $N_\downarrow=3$. Here we again observe the small-amplitude oscillations on top of an overall decrease of $\Delta$. These oscillations are due to the coupling to both the lowest dipole and the lowest breathing modes. The coupling to different collective modes due to the different spin population distribution may be regarded as a manifestation of spin-orbit coupling. 

\section{Conclusion} 
We have presented an exact mapping that maps a 1D spinor quantum gas of arbitrary spin with contact $s$-wave interaction to the direct product of a spinless Fermi gas interacting with a short-range $p$-wave pesudo-potential and a spin parity projection operator. This mapping allows us to construct straightforwardly the multi-branch spin chain model for strongly interacting spinor gases, using which we calculated the interaction-induced shift of the lowest breathing mode frequency, as well as the quench dynamics of a spin-1/2 Fermi gas. Our work demonstrates that the multi-branch spin chain model can accurately capture both the static and the dynamical properties of the system.  

From a conceptual point of view, the mapping allows us to gain new insights into the strongly interacting 1D systems, in particular, the interplay between the charge and the spin degrees of freedom. For example, the multi-branch spin chain model can be intuitively understood under the framework of first-order perturbation theory using the mapped system. Furthermore, as the mapping itself is exact, one may in principle take the perturbation calculation to higher orders in order to obtain more accurate results. From a technical point of view, this mapping significantly reduces the size of the effective Hilbert space. As a result, we can construct very efficient numerical tools to investigate the properties of the system. Our study will thus open up many avenues of research in the study of 1D systems.

\textit{Acknowledgment} --- We would like to thank Prof. Xiwen Guan for many insightful comments, and Nikolaj Zinner for his comment on an earlier version of the paper. This research is supported by US NSF and the Welch Foundation (Grant No.
C-1669).

\appendix
\section{Derivation of $H_p$ using a two-particle system}
In this Appendix, we use a two-particle system to derive the form of $H_p$ in Eq.~(\ref{pesudopotential}).
Let us consider a system consisting of two fermions governed by Hamiltonian
\begin{equation}
H=\sum_{i=1,2}[-\partial_i^2/2+V(x_i)]+g\delta(x_{12}) \,. \label{5}
\end{equation}
The non-trivial eigenstates must have even spatial wave function and, correspondingly, odd spin wave function. Due to $s$-wave interaction term in the form of a Dirac $\delta$-function, the relative spatial wave function $\psi(x_{12})$ satisfies the boundary condition:
\begin{equation}
\lim_{x_{12} \rightarrow 0^+} g\psi (x_{12}) =2\psi'(x_{12}) \,. \label{bc}
\end{equation}
Now consider the Hamiltonian $H_F=H_0+H_p$ with
\begin{eqnarray}
H_p=-\frac{4}{g}\overleftarrow{\partial}_{x_{12}}\delta(x_{12})\overrightarrow{\partial}_{x_{12}}\otimes  \hat{P}^a_{1} \,, \label{7}
\end{eqnarray}
where the single particle part acts on the spinless fermion space, and the interaction part decribes a $p$-wave interaction and contains the spin projection operator $\hat{P}^a_{1}$ which projects out the anti-symmetric spin states (if the original system consists of bosons, we should use the symmtric spin projection operator $\hat{P}^s_{1}$ instead). The relative spatial wave function of eigenstates of $H_F$, which we denote as $\psi_F(x_{12})$, must be an odd function of $x_{12}$. In fact, they can be constructed from the even eigenfunctions of Hamiltonian $H$ in Eq.~(\ref{5}) as follows:
\begin{equation}
\psi_F(x_{12}) ={\rm sgn}(x_{12}) \psi(x_{12}) \,,
\end{equation} 
It is straightforward to show \cite{Sen1999,Sen2003,Grosse2004,Hao2007,Girardeau2004_1,Girardeau2004_2} that $\psi_F$ satisfies the same boundary condition (\ref{bc}).

We can insert a sector function $\theta^1=\theta(x_2-x_1)$, together with a normalization factor $2!$ to the spatial part of the $p$-wave interaction term in Hamiltonian (\ref{7}):
\begin{equation}
H_p= -\frac{4\cdot 2!}{g}\overleftarrow{\partial}_{x_{12}}\delta(x_{12})\theta^1\overrightarrow{\partial}_{x_{12}}\otimes  \hat{P}^a_{1} \,. \label{9}
\end{equation}
Doing this is not of much relevance for two particles, since for two-particle systems, there are only two spatial sectors defined as $x_1 \le x_2$ and $x_1 \ge x_2$, and accordingly there is only one boundary at $x_1=x_2$ shared by the two sectors. However, the inclusion of the sector function is essential for generalization into more particles, as in this case, the boundaries of different spatial sectors are different \cite{sector1}, and our mapping is defined only in one spatial sector. Generalizing (\ref{9}) to an $N$-particle system, we can write down the $p$-wave pseudopotential term as 
\begin{equation}
H_{p}=- \frac{4N!}{g}\,\sum_{i=1}^{N-1}\overleftarrow{\partial}_{x_{i,i+1}}\delta(x_{i,i+1})\theta^1\overrightarrow{\partial}_{x_{i,i+1}}
\otimes \hat{P}^{s,a}_i \,. \label{pesudopotential1}
\end{equation}
By using $\partial_{x_{i,i+1}}=\frac{1}{2}\partial_i-\frac{1}{2}\partial_{i+1}$ together with the fact that the relative spatial wave function is odd, it is easy to show that Eq.~(\ref{pesudopotential1}) is equivalent to Eq.~(\ref{pesudopotential}) in the main text.

\section{SO(2,1) algebra for harmonic oscillator}
We use the same convention as in \cite{Nishida2007}. The generators for the center-of-mass (COM) harmonic oscillator algebra and the SO(2,1) algebra can be made of generators from Schr\"{o}dinger algebra, for which all the commutation relations are known \cite{Nishida2007}. The operators we use include
\begin{equation}
K=\int dx \,xn(x)\,, \;\;\;\; P=\int dx\,j(x)\,, \;\;\;\; D=\int dx\,xj(x)\,,  \label{Salgebra_1}
\end{equation}
\begin{equation}
H=-\frac{1}{2}\int dx\,\psi^{\dagger}(x)\partial^{2}\psi(x)\,, \;\;\;\; C=\int dx\,\frac{x^{2}}{2}n(x)\,, \label{Salgebra_2}
\end{equation}
where $j(x)=-\frac{i}{2}(\psi^{\dagger}(x)\partial\psi(x)-\partial\psi^{\dagger}(x)\psi(x))$ is the current density. Here $K$ represents the COM coordinate, $P$ the total momentum, $H$ the kinetic energy, $C$ the trapping potential, and $D$ the generator for scaling transformation. Again we have used the trap units with $\hbar=m=\omega=1$. We can define COM ladder operators (without normalization) $Q$ and $Q^\dag$, and COM Hamiltonian $H_0^{\rm c}$ as
\begin{align}
Q=\frac{K+iP}{\sqrt{2}} \,,\;\;\;\;\;
Q^\dag=\frac{K-iP}{\sqrt{2}} \,,\;\;\;\;\;
H^{\rm c}_0=\frac{\{Q,Q^\dag \}}{2N} \,.
\end{align}
These three operators form a harmonic oscillator algebra for the COM motion. 

The operators for the relative motion can be constructed as: 
\begin{align}
B&=\frac{1}{2}\left[H-C+iD\right]-\frac{Q^2}{2N} \,, \label{B}\\
B^{\dagger}&=\frac{1}{2}\left[H-C+iD\right]-\frac{Q^{\dagger2}}{2N} \,,\\
H_0^{\rm i}&=H+C-H_0^{\rm c}\,,
\end{align}
which form a closed SO(2,1) algebra as they obey the following
commutation relations:
\begin{align}
[H_0^{\rm i},B]=-2B\,,\;\;\;\;\;
[H_0^{\rm i},B^{\dagger}]=2B^{\dagger}\,,\;\;\;\;\;
[B,B^{\dagger}]=H_0^{\rm i} \,.
\end{align}
These three operators only act on the internal degree of freedom. As a result, the internal degrees of freedom can be classified into irreducible representations (IRs) of this SO(2,1) algebra. Each IR is a tower of states with energy level spacing $2$. Together with the COM harmonic algebra, we can construct several lowest excited manifolds by acting $Q^{\dagger}$ and $B^{\dagger}$ on the ground state manifold. The five lowest manifolds are listed in Table~\ref{manifolds}. 
\begin{table}[h]
\begin{tabular}{c|ccccc}
degeneracy & & charge states \\
\hline
  5 & $Q^{\dagger4}\ket{0}$ & $Q^{\dagger2}B^{\dagger}\ket{0}$ & $B^{\dagger2}\ket{0}$ & $Q^{\dagger}\ket{\alpha}$ & $\ket{\beta}$ \\
  3 & $Q^{\dagger3}\ket{0}$ & $Q^{\dagger}B^{\dagger}\ket{0}$ & $\ket{\alpha}$ \\
  2 & $Q^{\dagger2}\ket{0}$ & $B^{\dagger}\ket{0}$  \\
  1 & $Q^{\dagger}\ket{0}$  \\
  1 & $\ket{0}$  \\
\end{tabular}
\caption{Several lowest manifolds constructed by acting ladder operators $Q^{\dagger}$ and $B^{\dagger}$ on the ground state. $\ket{0}$ stands for the ground state. $\ket{\alpha}$ and $\ket{\beta}$ stand for the beginning states of other SO(2,1) towers.}\label{manifolds}
\end{table}

We know that the fourth manifold is three-fold degenerate, but from $Q^{\dagger}$ and $B^{\dagger}$ we can only construct two states. So we need to introduce a new state denoted as $\ket{\alpha}$ having the property $Q\ket{\alpha}=B\ket{\alpha}=0$. This is where another SO(2,1) IR tower begins. Similarly we need to introduce $\ket{\beta}$ for the fifth excited manifold. The basis of each manifold can be constructed in this way by acting $Q^{\dagger}$ and $B^{\dagger}$ on lower states and introducing beginning states for new towers. 

\begin{widetext}
\section{Derivation of recursion relation Eq.~(\ref{factor})}
When we want to write down the first order perturbation of $H_p$ in the $n^{\rm th}$ manifold, we may need to calculate matrix elements whose most general form is
\begin{equation}
\frac{\bra{\alpha_0,\chi}Q^pB^mH_pB^{\dagger n}Q^{\dagger q}\ket{\beta_0,\chi'}}{\sqrt{\bra{\alpha_0,\chi}Q^pB^mB^{\dagger m}Q^{\dagger p}\ket{\alpha_0,\chi}\bra{\beta_0,\chi'}Q^qB^nB^{\dagger n}Q^{\dagger q}\ket{\beta_0,\chi'}}} \label{matrixelements_1}
\end{equation}
where $\chi$ and $\chi'$ stands for two spin states. $\alpha_0$ and $\beta_0$ stands for two arbitrary beginning states of two SO(2,1) IR towers and also with no COM excitation. They are excited by $B^{\dagger}$ and $Q^{\dagger}$ to the energy of the manifold we are considering. Since we are doing first order perturbation, the relation of $E_{\alpha_0}+p+2m=E_{\beta_0}+q+2n$ must be hold. 

Equation (\ref{matrixelements_1}) can be simplified as follows. Since $Q$ commute with $H_p$ and $B$, Eq.~(\ref{matrixelements_1}) is only nonzero when $p=q$ and $Q$ operators in the numerator and denominator are canceled. So we only need to consider
\begin{equation}
H_{{\rm sc},\alpha_0 \beta_0,mn}=\frac{\bra{\alpha_0,\chi}B^mH_pB^{\dagger n}\ket{\beta_0,\chi'}}{\sqrt{\bra{\alpha_0,\chi}B^mB^{\dagger m}\ket{\alpha_0,\chi}\bra{\beta_0,\chi'}B^nB^{\dagger n}\ket{\beta_0,\chi'}}} \label{matrixelements_2}
\end{equation}
where $E_{\alpha_0}+2m=E_{\beta_0}+2n$ must be hold. It is possible that $\alpha\ne\beta$ and $H_{{\rm sc},\alpha\beta,mn}$ nonzero, under this case $H_p$ will couple different charge states. In the first four manifolds listed in Table~\ref{manifolds}, we do not need to consider this as it can be straightforwardly shown that matrix elements of $H_p$ between two different states within the same manifold all vanish. However, this is no longer true for the fifth and higher manifolds since $\bra{\beta}H_pB^{\dagger2}\ket{0}$ in general is nonzero. If we are only concerned with the first 4 manifolds, we can further simplify Eq.~(\ref{matrixelements_2}) to 
\begin{equation}
H_{{\rm sc},m}=\frac{\bra{0,\chi}B^mH_pB^{\dagger m}\ket{0,\chi'}}{\bra{0,\chi}B^mB^{\dagger m}\ket{0,\chi'}} \label{matrixelements_3}
\end{equation}
where $0$ stands for the charge ground state. Equation (\ref{matrixelements_3}) can be written into a recursion relation \cite{Moroz2012} by using the known commutation relations of operators in the Schr\"{o}dinger algebra \cite{Nishida2007}. Since all we use to derive this recursion relation is using commutation relations among the operators defined in Eqs.~(\ref{Salgebra_1}) and (\ref{Salgebra_2}), as well as $H_p$, and their action on the charge degree of freedom, in the following we suppress the spin states $\chi$ and $\chi'$. 
By switching $B$ and $H_p$ twice and denoting $\langle ... \rangle = \langle 0| ... |0 \rangle$, we can arrive at
\begin{align}
\begin{split}
H_{{\rm sc},m}	
&=\frac{\left\langle B^{m}H_{p}B^{\dagger m}\right\rangle }{\left\langle B^{m}B^{\dagger m}\right\rangle } \\
&=\frac{\left\langle B^{m-1}(H_{p}B+[B,H_{p}])B^{\dagger m}\right\rangle }{\left\langle B^{m}B^{\dagger m}\right\rangle } \\
&=H_{{\rm sc},m-1}+\frac{\left\langle B^{m-1}[B,H_{p}]B^{\dagger m}\right\rangle }{\left\langle B^{m}B^{\dagger m}\right\rangle } \\
&=H_{{\rm sc},m-1}+\frac{\left\langle B^{m-1}(B^{\dagger}[B,H_{p}]+[[B,H_{p}],B^{\dagger}])B^{\dagger m-1}\right\rangle }{\left\langle B^{m}B^{\dagger m}\right\rangle } \\
&=H_{{\rm sc},m-1}+\frac{\left\langle B^{m-1}B^{\dagger m-1}\right\rangle ^{2}}{\left\langle B^{m-2}B^{\dagger m-2}\right\rangle \left\langle B^{m}B^{\dagger m}\right\rangle }\left[H_{{\rm sc},m-1}-H_{{\rm sc},m-2}\right]+\frac{\left\langle B^{m-1}[[B,H_{p}],B^{\dagger}]B^{\dagger m-1}\right\rangle }{\left\langle B^{m}B^{\dagger m}\right\rangle } \label{recursion1}
\end{split}
\end{align}
Let us next consider the last term. Since $Q$ commutes with $H_p$ and $B$, we can ignore all the $Q$ parts in $B$ [see Eq.~(\ref{B})]. Consider the commutator $[B,H_{p}]$. First let us prove $[C,H_p]=0$. Since $H_p$ is in first quantized form, we also use the first quantized form of $C$, with which we have
\begin{equation}
[C,H_{p}]=\left[\sum_{i=1}^{N}\frac{1}{2}x_{i}^{2},-\frac{4N!}{g}\sum_{i=1}^{N-1}\overleftarrow{\partial}_{i}\delta(x_{i}-x_{i+1})\theta^{1}\overrightarrow{\partial}_{i}\otimes\hat{P}_{i}^{s,a}\right]
\end{equation}
In this expression, note that each term in the spatial part of $H_p$ only acts on relative coordinate $x_{ii+1}=x_i-x_{i+1}$, and $C$ can also be separated into one part containing relative coordinates $x_{ii+1}$ and another part containing the COM coordinate $\sum x_i/N$. The nonzero contribution can only come from the commutator
\begin{equation}
\left[x^{2},\overleftarrow{\partial}\delta(x)\theta(x)\overrightarrow{\partial}\right]=2x\delta(x) \theta(x)\overrightarrow{\partial}-\overleftarrow{\partial}\delta(x)\theta(x)2x \,,
\end{equation}
which is 0 since $x\delta(x)=0$. Here we have also ignored the regularization point splitting in $\partial_{-}$, since in the derivation of the recursion formula, we only need to consider continuous wavefunctions. Therefore we have proved that 
\begin{equation}
[C,H_{p}]=0\,. \label{commC}
\end{equation}
Next we consider $H_p$'s scaling dimension. Since $H_p$ is made of two spatial derivatives and a delta function (also a $\theta$ function whose scaling dimension is 0), it has scaling dimension 3, which means 
\begin{equation}
[D, H_p]=i\Delta_{H_p}H_p,\;\;\Delta_{H_p}=3 \,. \label{commD}
\end{equation}
Using Eqs.~(\ref{commC}) and (\ref{commD}), the commutator $[B,H_{p}]$ can be written as
\begin{equation}
[B,H_{p}]=\frac{1}{2}\left[H,H_{p}\right]+\frac{1}{2}\Delta_{H_{p}}H_{p} \,.
\end{equation}
Now consider its commutation relation with $B^{\dagger}$
\begin{align}
\begin{split}
\left[\left[B,H_{p}\right],B^{\dagger}\right]
&=\frac{1}{4}\left[\left[H,H_{p}\right]+\Delta_{H_{p}}H_{p},H-C+iD\right] \\
&=\frac{1}{4}\left[\left[H,H_{p}\right]+\Delta_{H_{p}}H_{p},H_{\rm osc}\right]-\frac{1}{2}\left[\left[H,H_{p}\right],C\right]+\frac{1}{4}\left[\left[H,H_{p}\right],iD\right] +\frac{1}{4}\Delta_{H_{p}}^{2}H_{p}\,. \label{comm2}
\end{split}
\end{align}
In the second step we have used $[C,H_p]=0$ and $[D,H_p]=i\Delta_{H_p}H_p$. And we introduce $H_{\rm osc}=H+C$ which is the harmonic oscillator Hamiltonian. Using Jacobi identity followed by the commutation relations $\left[C,H\right]=iD$, $\left[D,H\right]=2iH$, $[C,H_p]=0$ and $[D,H_p]=i\Delta_{H_p}H_p$, the second and third terms of Eq.~(\ref{comm2}) can be written as
\begin{align}
\begin{split}
\left[\left[H,H_{p}\right],C\right]&=-\left[\left[H_{p},C\right],H\right] -\left[\left[C,H\right],H_{p}\right]=-\left[iD,H_{p}\right]=\Delta_{H_p}H_p \,, \\
\left[\left[H,H_{p}\right],iD\right]&=-\left[\left[H_{p},iD\right],H\right]- \left[\left[iD,H\right],H_{p}\right]=-\left[\Delta_{H_{p}}H_{p},H\right]+\left[2H,H_{p}\right]=-\left[\Delta_{H_{p}}H_{p},H_{\rm osc}\right]+\left[2H_{\rm osc},H_{p}\right]\,.
\end{split}
\end{align}
Since we are going to calculate $\bra{0} B^{m-1}[[B,H_{p}],B^{\dagger}]B^{\dagger m-1}\ket{0}$ and $B^{\dagger m-1}\ket{0}$ is an eigenstate of $H_{\rm osc}$, all the terms which are commutators with $H_{\rm osc}$ vainsh. Only the second and forth term in Eq.~(\ref{comm2}) remain. Finally the last term in Eq.~(\ref{recursion1}) can be written as
\begin{equation}
\frac{\left\langle B^{m-1}[[B,H_{p}],B^{\dagger}]B^{\dagger m-1}\right\rangle }{\left\langle B^{m}B^{\dagger m}\right\rangle }=\frac{1}{4}\Delta_{H_{p}}(\Delta_{H_{p}}-2)\frac{\left\langle B^{m-1}B^{\dagger m-1}\right\rangle }{\left\langle B^{m}B^{\dagger m}\right\rangle }H_{{\rm sc},m-1} \label{thirdterm}
\end{equation}
The normalization can also be easily evaluated by using $[B,B^{\dagger}]=H_0^{\rm i}$, from which we have
\begin{equation}
S_{m}=\frac{\left\langle B^{m+1}B^{\dagger m+1}\right\rangle }{\left\langle B^{m}B^{\dagger m}\right\rangle }=(m+1)(m+E_{0}^{\rm i}) \,.\label{normrecursion}
\end{equation}
where $E_0^{\rm i}=N^2/2-1/2$ is the ground state internal energy. Putting Eqs.~(\ref{recursion1}), (\ref{thirdterm}), and (\ref{normrecursion}) together, a recursion relation is obtained
\begin{equation}
H_{{\rm sc},m}-H_{{\rm sc},m-1}=\frac{(m-1)(m-2+E_{0}^{{\rm i}})}{m(m-1+E_{0}^{{\rm i}})}\left[H_{{\rm sc},m-1}-H_{{\rm sc},m-2}\right]+\frac{1}{4}\frac{\Delta_{H_{p}}(\Delta_{H_{p}}-2)}{m(m-1+E_{0}^{{\rm i}})}H_{{\rm sc},m-1} \,. \label{recursion2}
\end{equation}
The spin chain Hamiltonian for the first excited manifold and for the dipole state of the second excited manifold, which correspond to $m=0$, are the same as ground state spin chain Hamiltonian (when a constant shift is neglected):
\begin{equation}
H_{{\rm sc}}^{(1)}=H_{{\rm sc}}^{(Q)}=H_{{\rm sc},0}=H_{{\rm sc}}^{(0)}\,.
\end{equation}
The spin chain Hamiltonian for breathing state of the second excited manifold, which corresponds to $m=1$, is proportional to the ground state spin chain Hamiltonian:
\begin{align}
H_{{\rm sc}}^{(B)}=H_{{\rm sc},1}=\left[1+ \frac{3}{2(N^2-1)}\right]H_{{\rm sc},0}=\left[1+ \frac{3}{2(N^2-1)}\right]H_{{\rm sc}}^{(0)}\,.
\end{align}
We stress here that we cannot obtain a similar recursion formula like Eq.~(\ref{recursion2}) for the most general matrix elements $H_{{\rm sc},\alpha_0 \beta_0,mn}$ (Eq.(\ref{matrixelements_2})) for arbitrary $\alpha_0$ and $\beta_0$ case. This is because $H_p$ may couple different charge states within the same manifold, resulting in an entanglement between the spatial and the spin sectors.

\end{widetext}


\begin{thebibliography}{99}
\bibitem{regularization} The minus sign in $x_{ij}^{-}$ means $x_{ij}^{-}=x_i-x_j-0^{+}$, which is needed for regularization as can be seen in Ref.~\cite{Girardeau2004_1,Girardeau2004_2}.
\bibitem{note} Here we focus on the case with repulsive interaction, i.e., $g>0$, although the spin chain model is also valid for the upper branch in the strongly attractive regime.
\bibitem{sector1} For example, in the case with 3 particles. The spatial sector $x_1 \le x_2 \le x_3$ has two boundaries defined by $x_1=x_2<x_3$ and $x_1 < x_2=x_3$; while the spatial sector $x_1 \le x_3 \le x_2$ has two boundaries defined by $x_1=x_3<x_2$ and $x_1 < x_3=x_2$. Hence these two spatial sectors share only one common boundary.
\end{thebibliography}
\end{document}